# Quantum theory methods as a possible alternative for the double-blind gold standard of evidence-based medicine: Outlining a new research program

Diederik Aerts[1], Lester Beltran[1], Suzette Geriente[1], Massimiliano Sassoli de Bianchi[1,2], Sandro Sozzo[3], Rembrandt Van Sprundel[1] and Tomas Veloz[1,4]

(1) Center Leo Apostel for Interdisciplinary Studies, Brussels Free University, Krijgskunde- straat 33, 1160 Brussels, Belgium. E-mails: diraerts@vub.ac.be, lestercc21@yahoo.com, sgeriente83@yahoo.com, msassoli@vub.ac.be

(2) Laboratorio di Autoricerca di Base, 6917 Lugano, Switzerland. E-mail: autoricerca@gmail.com

(3) School of Business and Centre IQSCS, University Road, LE1 7RH Leicester, United Kingdom. E-mail: ss831@le.ac.uk

(4) Universidad Andres Bello, Departamento Ciencias Biológicas, Facultad Ciencias de la Vida, 8370146 Santiago, Chile and Instituto de Filosofía y Ciencias de la Complejidad, Los Alerces 3024, Ñuñoa, Chile. Email: tveloz@gmail.com

**Abstract.** We motivate the possibility of using notions and methods derived from quantum physics, and more specifically from the research field known as 'quantum cognition', to optimally model different situations in the field of medicine, its decision-making processes and ensuing practices, particularly in relation to chronic and rare diseases. This also as a way to devise alternative approaches to the generally adopted double-blind gold standard.

Human decision and its ensuing practices play an important role in many domains of human endeavor. When decisions are taken such that their consequences are determinative for the health and care of sick people, an element of even greater importance is added. However, if compared to decision making in economics, psychology, cognitive science, or finance, medical decision making has not yet concentrated very much on structured decision model building [1, 2, 3, 4][1]. Several reasons for this can be identified. First of all, decisions in medicine touch upon some of the most delicate and valued aspects of us as human beings, possibly also related to matters of life and death, and for this a structured theoretical modeling of their processes might seem almost impossible [5, 6, 7, 8, 9]. Secondly, very complicated ethical questions are connected with even simple decisions and their subsequent practices in medicine, which also seems to indicate that only highly contextual and ad hoc reasoning is at place [10, 11, 12]. Thirdly, and more generally, the concrete situations appearing with respect to decision-making and its associated practices in medicine confront aspects that are intrinsically extremely complex, not yet well understood, and possibly also paradoxical, when viewed from a certain perspective. We think here, for example, of the ethical issues arising with the use of placebo controls in clinical trials [10], and other types of problems, of different nature, related to clinical research in general [11, 12]. We also think of those decisions taking into account several clinical trials testing the same illness, which often do not converge because of the inherent complex unpredictability [12]. And, finally, we think of the placebo effect as a general phenomenon, indicating that there are aspects of the medical practice that can themselves influence in a not yet understood way the state of health or illness of the patients who are subjected to them [13, 14, 15, 16].

---

[1] Although we mention that the 'Society for Medical Decision Making' (http://smdm.org), with its peer reviewed journal, is an important forum for many of the delicate aspects related to medical decision making.

The above-mentioned complexities mask in some way the deeper structure of decision-making and its ensuing practices in medicine, so that it seems difficult to conceive of a theory general enough to provide a realistic modeling. We believe, however, that such a theory is possible, and that a version of it already exists, applied to other domains such as economics, finance, psychology and cognitive science. It is called 'quantum cognition' and its structure is inspired by the mathematical models of quantum theory [17, 18, 19, 20, 21, 22]. That quantum structure can also play a role in the modeling of human decision-making in medicine should however not come as a surprise. Indeed, we know that one of the major fingerprints of quantum theory is the presence of an intrinsic and irreducible influence of the measurement process on the state of the entity under study, often referred to as an 'observer effect' [23]. If we think again of the general phenomenon of the placebo effect, it incorporates exactly this type of influence of the medical practice on the state of health or illness of the patient. As an example, we can mention the not commonly known fact that even when patients are informed that they are being treated with a placebo, an effect still remains present [16].

Another phenomenon more directly related to the thought processes taking place during medical decisions, which has been extensively studied in quantum cognition, is the so-called 'conjunction fallacy'. It consists in estimating the probability of the joint occurrence of two events to be higher than the probability of occurrence of only one of the two events [24]. The effect is called 'double conjunction fallacy' if the probability of joint occurrence of the two events is estimated to be higher than that of both individual events. In classical probability theory, which is based on Boolean logic, the probability of occurrence of the conjunction of two events is always smaller or equal to the probability of occurrence of only one of the two events, which is the reason why the phenomenon is indicated as a 'fallacy', meaning an 'error in reasoning' [24].

In the original article where the phenomenon was identified, an example of a conjunction fallacy occurring in medicine was already given [24]. More specifically, it was observed that internists responded that a 55-year-old woman was more likely to experience, following a blockage of an artery in the lungs (pulmonary embolism), the conjunction of 'shortness of breath (dyspnea) and weakness in one side of the body (hemiparesis)', than 'weakness in one side of the body'. A simple reasoning using elementary logic shows why the conjunction fallacy's phenomenon is a fallacy of classical probability theory. Indeed, whenever the woman in question has 'shortness of breath and weakness in one side of the body', she obviously also has 'weakness in one side of the body'. Hence, the probability of occurrence of the former must always be smaller or equal to the probability of occurrence of the latter. Why then people do make abundantly this error, leading to the phenomenon of the conjunction fallacy? Quantum cognition offers a new perspective on this question. Indeed, in addition to being a new physical theory, quantum mechanics also contains a new non-classical probability model, and within the structure of the quantum probability model the phenomenon of the conjunction fallacy can occur without it being considered to be an error, but a direct consequence of its particular structure.

In our Brussels approach to quantum cognition, we have studied in great detail a phenomenon very similar to the conjunction fallacy, which is referred to as the 'pet-fish problem', or the 'guppy effect'. It was identified in psychology in the domain of 'concept research' even before the conjunction fallacy was introduced [25]. It consists in observing that 'guppy', or 'goldfish', are considered to be more typical of a 'pet-fish' than of only a 'pet', or only a 'fish'. Hence, the feature of 'typicality' behaves weirdly with respect to the conjunction, in the sense that the conjunction 'pet-fish' rates higher in typicality than the single components 'pet' and 'fish', for specific entities such as a 'guppy' or a 'gold-fish'. In our Brussels group, we have worked out quantum models for the guppy effect in great detail, and we believe that it is an archetype for what happens when the conjunction fallacy manifests in experimental situations, leading to a much better understanding of the phenomenon than in traditional explanations, where it is merely classified as a probability judgment error [26, 27, 28, 29]. This also because, a much less known probabilistic variant of the original pet-fish problem, using 'membership weights' instead of 'typicalities', has also been abundantly tested, giving rise to analogous behaviors for the conjunction [30, 31, 32].

It is by means of the above-mentioned modeling techniques, which we developed to address the conjunction fallacy and the pet-fish problem in human cognition, that we want to investigate similar effects that can appear in the medical practice. We observed that the conjunction fallacy has been studied very little in relation to medical thought, though considering the importance of medical decisions one would have imagined differently. The pet-fish variant of it has not yet been considered at all. We found one article where it is shown that half of early medical students estimated the probability that a patient with a common cold would experience the symptoms of 'runny nose and diarrhea' to be higher than that of the patient experiencing the symptom of 'diarrhea' [33]. However, we also found a very recent and thorough investigation where an impressive amount of data was obtained, showing the occurrence of the double conjunction fallacy in a systematic way in medicine [34]. The data collected in this work interests us particularly, because, as the authors themselves recognize, there are actually no convincing models of decision-making for the double conjunction fallacy [35, 36, 37, 38, 39]. This is the reason why the latter is often believed to only appear very sporadically, hence not being representative of the mechanism at play in a human mind when a conjunction fallacy occurs. The authors of the above-mentioned study disagree with this view and claim that their experiments show that double conjunction fallacies are no exception and appear abundantly whenever some specific cognitive situations arise. What is interesting is that in the quantum model of the pet-fish situation that we developed in our group [17, 26, 27, 28, 29], the single and double conjunction fallacies can be modeled in a very natural way, and result both from a same fundamental mechanism, which we plan to use also as an explanation for, and to further the analysis of, the situations put forward in [34].

Having said that, we would like now to mention a situation where the application of quantum statistics – the probabilistic formalism of quantum mechanics – would be of specific value in possibly providing an alternative to the double-blind gold standard of evidence-based medicine [2]. Chronic diseases, in particular if they are only vaguely diagnosed, pose a special challenge to the gold standard of evidence-based medicine. Indeed, the ethical and technical aspects that are related to its corner stone, which is the final placebo-controlled trial, can many times pose almost insurmountable obstacles. How can one treat a whole group of chronically ill patients, in the long term, in a placebo well controlled trial, without doing some harm to the placebo group [40]? Also, chronic diseases lend themselves better than any other category of diseases to statistical analysis, considering that they evolve over a long period of time. A thorough study of data collected from people who suffer from chronic disease would therefore be the perfect starting point for applying quantum statistics to medicine.

There are other aspects that make chronic diseases particularly suitable for a deep analysis by means of quantum statistics, as an alternative to the double-blind gold standard of evidence-based medicine. Patients who suffer from a chronic disease often have a lot of diversified data to submit, since they usually have tried out several therapies, depending on the ups and downs of their health conditions. These patients are often also prepared to actively participate in the production of an even wider variety of data concerning their state of health. Also, they are often actively engaged in their disease processes, have read about them, searched information on the World-Wide Web, participated in patients' forums, etc. All these are aspects that favor a statistical investigation, at least if the statistical tools used are able to take into account all these self-induced aspects of the patients as a consequence of their active attitude towards different aspects of their diseases. This ability of managing self-induced aspects in a way that the results obtained from the statistics are still reliable is exactly one of the salient characteristics and highlights of quantum statistics.

There is a second category of diseases which we would like to also take into account, namely the so-called 'rare diseases' and/or 'orphan diseases'. For different reasons than the chronic ones, they also pose special challenges with respect to the gold standard of scientific evidence in medical research, which consists of having a medicament pass the series of type 0, I, II, III and IV trials on human subjects, ending in a double-blind placebo-controlled experiment [41].

---

[2] The term 'gold standard' usually refers to a (randomized) double-blind clinical trial, the latter being an experiment that has been planned and designed to test the efficacy/effectiveness of a treatment, by observing its effects in a group of patients, by comparison with those observed in a similar group of patients only receiving a control (placebo) treatment. Blinding refers to the masking of information about the test, in order to reduce or eliminate possible biases (e.g., preferences, expectations), and a double-blinding corresponds to the situation where both the patients and the investigators are blind.

Indeed, after a medicine has succeeded in passing the initial research phases of development in vitro and the trials on animals, subsequent trials on humans need to take place in a hospital, instead of a research laboratory environment, and therefore are much more expensive than the trials of the preceding phases. For this reason, they usually fall outside the financial reach of research groups that are often located at universities, but they also fall outside of the interest of the pharmaceutical companies, which for non-rare-diseases will instead customarily finance these extra costs for the trials on humans, considering the earning prospects in sales of the tested medicament. For orphan diseases, this 'gold standard testing' of new medicaments frequently breaks down, because the turnover that can be generated is too meager.

Another challenge for rare diseases, which in some cases can even be more serious, is to find a sufficiently large group of patients who can participate in a gold standard approach. As a consequence, very often the development of new medicaments for rare diseases stops even before they are tested on human subjects [42, 43]. As a result of the sporadic nature of most of these diseases, a direct application of standard statistical methods is clearly not straightforward, but there are reasons to believe that more could be learned about them by studying how to optimally use quantum statistical methods in these less favorable situations, particularly once the full potential of quantum statistics has become better understood in applications to chronic diseases. In other words, rare/orphan diseases, due to their intrinsic limitations with respect to the double-blind gold standard, could profit from quantum statistical methods, although this can realistically happen only in a later stage, after more experience and results have been gathered with the more favorable situation of chronic diseases.

Coming back to the latter, on top of the complications of long-term decision-making in relation to the question of 'how to reach the best possible medical practice', with respect to chronic diseases one is also confronted with the strange effect called 'the disability paradox'. More specifically, many people with serious and persistent disabilities report that they experience a good or excellent quality of life, whereas to most external observers these individuals seem to live an undesirable daily existence [44, 45, 46, 47, 48]. This disability paradox shows how the phenomenon of chronic disease, as a consequence of its deep and long-standing influence on the life of an individual, confronts the 'quality of life state' with the 'normality of life state', and the fact that they are not a priori equal.

Up to here, we have considered quantum cognition mainly with respect to the new quantum statistics it incorporates, but it additionally provides a new non-classical paradigm for the conceptual and cognitive world, being able to model contexts and their effects on the state of an entity in a very refined and powerful way. Our hypothesis is that a quantum cognition approach, and more specifically the operational-realistic formalism developed in our Brussels' group [49, 50], can shed new light and powerful explanations on paradoxical situations like the disability paradox, or the placebo effects, and more generally on the phenomenon of disease processes which themselves can influence (and can be influenced) by the state of the patients, for instance as regards the perceived quality of their life. Indeed, next to providing a non-classical probability model giving rise to a quantum statistics, quantum theory provides, as we said, a powerful scheme to model contexts and their influence on the states of the entities that are subjected to them. It is also this aspect of quantum cognition that we plan to explore in our study of medical situations like those described in the disability paradox.

More generally, and on a theoretical level, notions such as 'context', 'emergence' and 'interference', none of which can be modeled by approaches inspired by classical theories (such as 'fuzzy sets theory', for example), can instead be addressed by the existing quantum cognition formalisms, which is the reason why very complex situations, apparently only allowing ad hoc models, can be approached at a deeper foundational level [17, 18, 19, 20, 21, 49, 50]. We are thus confident that advantages similar to those encountered in economics, finance, psychology and cognitive science will also be revealed if these new formalisms are applied to the domain of medicine, to investigate situations like those mentioned above, and possibly many others as well.

On a practical level, guided by the theoretical structure of these quantum cognition formalisms, we plan to develop and then use the statistical models they subtend to a point where direct studies of individual cases, for example of chronically ill patients, could become complementary to those based on the gold standard of evidence-based medicine, and even substitute them in those situations where they are almost not applicable, for ethical, technical or financial reasons. Since the quantum statistics derived from quantum cognition models are in principle capable of also incorporating the effects of the medical treatment on the state of health or illness at a fundamental level, there is indeed the possibility to develop a completely new way to take into account the placebo effect, such that quantum statistical analysis of large enough data might become a substitute for the double-blind placebo trials, allowing for results of possibly equal, if not even superior, scientific quality.

To conclude, we plan to apply quantum inspired methods to a variety of medical situations, to extract meaningful information about them and possibly also better understand the mechanisms underlying them, starting for this from the large sets of data about chronic diseases and the patients that are affected by them, then continuing, possibly, with the much smaller sets of data about rare diseases. And of course, in case the research program we have here outlined – in which we encourage the community of quantum cognitivist scientists to participate – will prove to be successful, its benefits will also be for the non-rare and non-chronic diseases, by possibly reducing the costs of the double-blind tests also for them, with the additional beneficial effect that smaller companies will be able to better compete and propose innovative medicaments, in a pharmaceutical marketplace which is today dangerously dominated by the multinational corporations.